\documentclass[fleqn,oneside]{article}
\usepackage{espcrc2}

\usepackage{graphicx}
\usepackage{epsfig}

\newcommand{\AmS}{{\protect\the\textfont2
  A\kern-.1667em\lower.5ex\hbox{M}\kern-.125emS}}

\title{Stacked Josephson junction SQUID}

\author{V.M.Krasnov\thanks{{\it E-mail address}:
krasnov@fy.chalmers.se}
\newline
\addressmark[]{Department of Microelectronics and Nanoscience, Chalmers
University of Technology, S-41296 G\"oteborg, Sweden}
 }

\begin{document}

\begin{abstract}
Operation of a Superconducting Quantum Interference Device (SQUID)
made of stacked Josephson junctions is analyzed numerically for a
variety of junction parameters. Due to a magnetic coupling of
junctions in the stack, such a SQUID has certain advantages as
compared to an uncoupled multi-junction SQUID. Namely,
metastability of current-flux modulation can be reduced and a
voltage-flux modulation can be improved if junctions in the stack
are phase-locked. Optimum operation of the SQUID is expected for
moderately long, strongly coupled stacked Josephson junctions. A
possibility of making a stacked Josephson junction SQUID based on
intrinsic Josephson junctions in high-$T_c$ superconductor is
discussed. \\



\end{abstract}

\maketitle

\section{Introduction}

Layered high-$T_c$ superconductors (HTSC) represent natural stacks
of atomic scale intrinsic Josephson junctions (IJJ's)
\cite{Kleiner1}. Indeed, an interlayer spacing of 15.5 $\AA$ was
estimated from a periodic Fraunhofer modulation of Fiske steps in
Bi-2212 mesas \cite{Fiske}. At present, high quality SIS-type
IJJ's can be fabricated using HTSC mesa structures
\cite{KrasnovT,KrasnovH}. IJJ's with their record large $I_c R_n$
values $\sim 10 mV$ are attractive candidates for cryoelectronics.

Here I analyze a possibility of making a SQUID, based on IJJ's.
Because the interlayer spacing in HTSC is very small, it is likely
that such a SQUID will contain several stacked Josephson junctions
(SJJ's), as shown schematically in Fig. 1. From previous studies
it is known that operation of a multi-junction SQUID, consisting
of series connected uncoupled junctions, is problematic due to the
existence of metastable states and a small current-flux modulation
\cite{Lewand,Darula,Konopka}. The main difference between series
connected and stacked junctions is that SJJ's are coupled with
each other, which may result in qualitatively different behavior
of SJJ's, see eg. recent review \cite{Review}.

The main question which is raised in this paper, is whether the
operation of a multi-junction SQUID can be improved by coupling
and phase-locking in stacked Josephson junctions.

\begin{figure}[ht]
\begin{minipage}{0.45\textwidth}
\epsfxsize=0.73\hsize \centerline{ \epsfbox{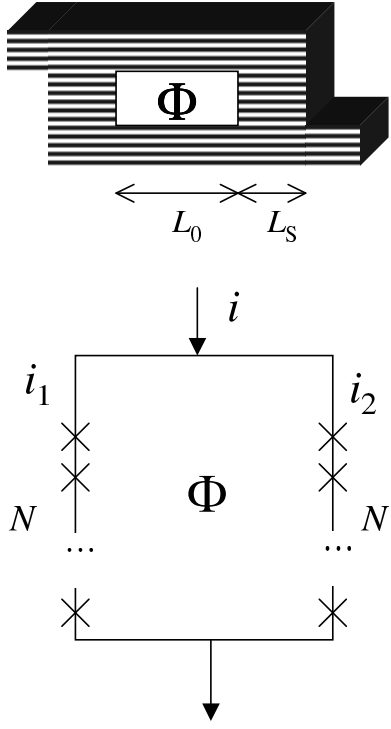} }
\end{minipage}
\vspace{-12pt} \caption{A schematic view and a diagram of a
stacked Josephson junction SQUID.} \label{autonum1}
\end{figure}

\section{Magnetic coupling in stacked junctions}

I will consider a magnetically (inductively) coupled stack of $N$
junctions. The coupling is achieved via common electrodes and is
strong when they are thinner than the London penetration depth,
$\lambda_S$ (which is certainly the case for HTSC IJJ's).
Properties of magnetically coupled SJJ's are described by a
coupled sine-Gordon equation \cite{SBP},

\begin{equation}
\varphi_i^{\prime \prime }={\bf A}\cdot J_i sin(\varphi_i)-{\bf
J}_b. \label{Eq.1}
\end{equation}

\noindent Here $\varphi_i$ are gauge invariant phase differences,
'primes' denote in-plane spatial derivatives, ${\bf A}$ is a
symmetric tridiagonal $N\times N$ matrix with nonzero elements:
$A_{i,i-1}=-S_i/\Lambda _l$; $A_{i,i}=\Lambda _i/\Lambda _l$;
$A_{i,i+1}=-S_{i+1}/\Lambda _l$, where $\Lambda _i=t_i+\lambda
_{Si}coth\left( \frac{d_i}{\lambda _{Si}}\right) +\lambda
_{Si+1}coth\left( \frac{d_{i+1}}{\lambda _{Si+1}}\right) $,
$S_i=\lambda _{Si}cosech\left( \frac{d_i}{\lambda _{Si}}\right) $,
$d_i$ and $t_i$ are thicknesses of superconducting electrodes and
insulating barriers, respectively, and ${\bf J}_b$ is a bias term
\cite{Modes}.

The coupling strength $S$ is determined by off-diagonal elements
of ${\bf A}$. For identical junctions,

\begin{equation}
S \simeq 2ch^{-1}(d/\lambda_S). \label{Eq.2}
\end{equation}

\section{Short junctions, $L_S<\lambda_J$}

According to Eq.(1), SJJ's are coupled via the second derivative
of phase. The characteristic length for the phase variation is
given by the Josephson penetration depth, $\lambda _{J}=
\sqrt{\frac{\Phi _0c}{8\pi ^2J_{c}\Lambda _S}}$ \cite{Review}. For
short SJJ's, $L_S<\lambda_J$, magnetic coupling is inefficient and
a SJJ SQUID is described by similar expressions as an uncoupled
multi-junction SQUID, see the diagram in Fig. 1.

\begin{eqnarray}
i=i_1+i_2; \\ J_{1i}sin(\varphi_{1i})=i_1, \
J_{2j}sin(\varphi_{2j})=i-i_1; \\ \nonumber \varphi_{1i}=arcsin
\left(\frac{J_{1k}}{J_{1i}}sin(\varphi_{1k}) \right) + 2\pi n, \\
\varphi_{2j}=arcsin \left(\frac{i-J_{1k}
sin(\varphi_{1k})}{J_{2j}}\right) + 2\pi m;
\\ \Phi=
\frac{\Phi_0}{2\pi}\left[\sum{\varphi_{1i}}-\sum{\varphi_{2j}}
\right].
\end{eqnarray}

\noindent Here subscripts 1 and 2 correspond to left and right
legs of the SQUID, respectively, $\Phi = \Phi_e - i_1L_1+i_2L_2 $
is the total magnetic flux in the SQUID loop, $\Phi_e$ is an
external flux and $L_{1,2}$ are inductances of the SQUID legs. All
variables are expressed in terms of the phase difference in the
weakest junction "1k".

The critical current vs. flux characteristic of the SQUID with
$N=3$ uncoupled ($L_S=0$) junctions per leg with equal $J_c$'s is
shown in Fig. 2 by thin lines. It is seen that $I_c(\Phi)$
consists of a number of lobes. Each lobe has a width $N\Phi_0$
(not $\Phi_0$ as in a conventional SQUID). This can be easily
understood from Eq.(6): due to a summation in the right-hand side,
the overall flux increases $N-$fold as compared to a conventional
SQUID with a single junction per leg. Nevertheless, the overall
periodicity of the $I_c(\Phi)$ pattern is one flux quantum
$\Phi_0$. This is caused by the appearance of equivalent lobes
shifted by $\Phi_0$, corresponding to $2\pi$ shift of phase in the
junctions, see Eq.(6).

From such a simple analysis we can see two main disadvantages of
the uncoupled multi-junction SQUID: (i) The SQUID possesses
several metastable states and $I_c(\Phi)$ is not unique, (ii) The
overall modulation of $I_c(\Phi)$ (envelope of all lobes) strongly
decreases due to a mismatch between the width of the individual
lobe ($N\Phi_0$) and the spacing between lobes ($\Phi_0$).

\section{Long junctions, $L_S>\lambda_J$}

Magnetic coupling is effective in long stacks. However, as far as
SQUID application is concerned, the stack should not be too long
otherwise flux penetration inside the stack becomes complicated
due to the appearance of multiple quasi-equilibrium fluxon modes
\cite{Modes,Compar,Review}. The optimum stack should have a
moderate length, $L_S \sim 4 \lambda_J$, so that magnetic coupling
is effective, but fluxon modes in SJJ's are not pronounced.

\begin{figure}[ht]
\begin{minipage}{0.49\textwidth}
\epsfxsize=0.73\hsize \centerline{ \epsfbox{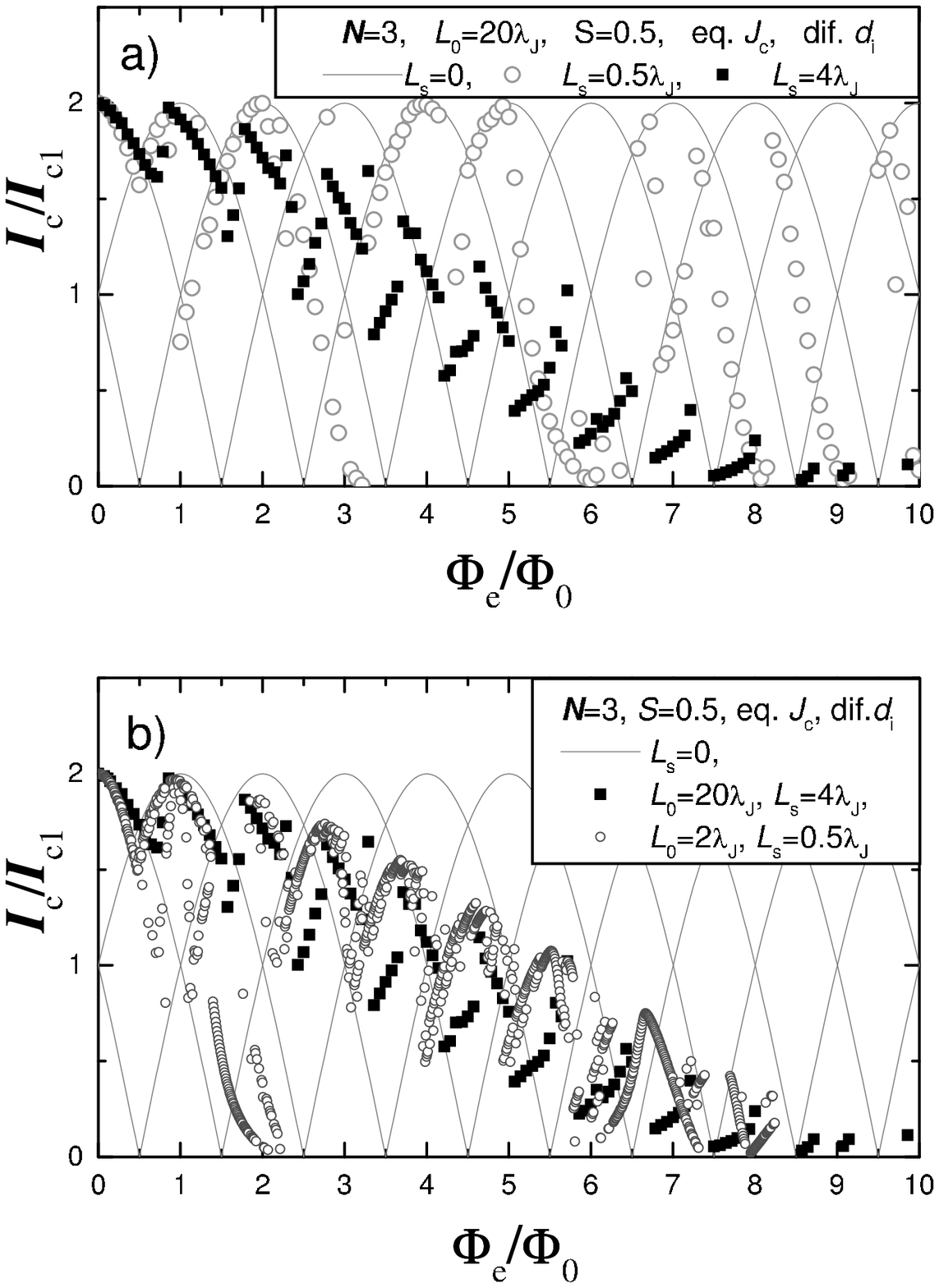} }
\end{minipage}
\vspace{-12pt} \caption{Simulated $I_c(\Phi_e)$ dependence of a
SQUID based on $N=3$ strongly coupled SJJ's  for different
junction lengths. It is seen that metastability of $I_c(\Phi_e)$
is reduced in moderately long SJJ's SQUID due to magnetic coupling
of junctions. } \label{autonum1}
\end{figure}

Static current-flux characteristics of a SJJ SQUID with the
geometry shown in Fig.1 were simulated numerically using a finite
difference method with successive interactions. For each stack,
Eq. (1) was solved together with Eqs.(3,6) and with boundary
conditions that magnetic induction is equal to external field on
the outer edges and to the field inside the SQUID loop for the
inner edges of SJJ's. Simulations were made for $N=3$ SJJ's in
each SQUID leg. The outer electrodes were assumed to be thicker,
$d_{1,4}>d_{2,3}$, in order to model bulk top and bottom bars of
the SQUID loop, see Fig. 1.

\begin{figure}[ht]
\begin{minipage}{0.49\textwidth}
\epsfxsize=0.73\hsize \centerline{ \epsfbox{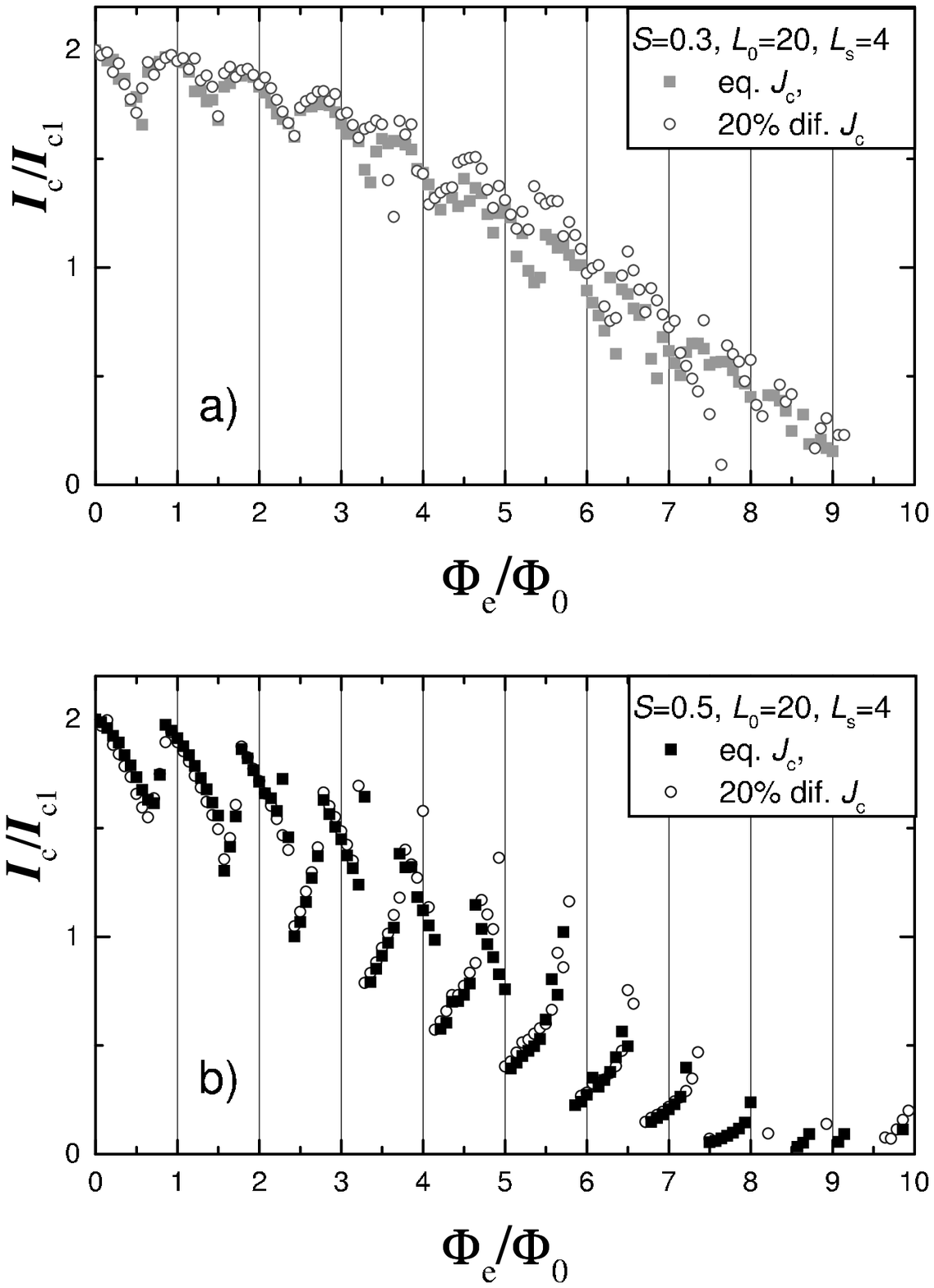} }
\end{minipage}
\vspace{-12pt} \caption{Simulated $I_c(\Phi_e)$ dependence of a
SQUID based on $N=3$ SJJ's for a) moderately coupled and b)
strongly coupled SJJ's. It is seen that strongly coupled SJJ's
SQUID becomes more stable and less susceptible to the spread in
junction parameters. } \label{autonum1}
\end{figure}

First we analyze the dependence of current-flux modulation on the
in-plane length of the stack, $L_S$. Fig. 2 shows simulated
$I_c(\Phi_e)$ patterns for a SQUID with strongly coupled SJJ's, $S
\simeq 0.5$, for different $L_S$. Parameters of SJJ's are:
$d_1=0.2\lambda_J$, $d_{2,3}=0.01\lambda_J$, $d_4=0.4\lambda_J$,
$t_{1-4}=0.01\lambda_J$, $\lambda_S=0.1\lambda_J$, critical
current densities of all junctions are identical. In Fig. 2 a)
$I_c(\Phi_e)$ patterns are shown for a fixed SQUID loop length,
$L_0=20\lambda_J$, for short SJJ's $L_S=0.5 \lambda_J$ (open
circles) and moderately long SJJ's $L_S=4 \lambda_J$ (solid
squares). For comparison, $I_c(\Phi_e)$ for the uncoupled
multi-junction SQUID ($L_S=0$) is shown by thin lines. It is seem
that $I_c(\Phi_e)$ of the SQUID with short SJJ's exhibit similar
metastability as the uncoupled multi-junction SQUID due to
inefficient magnetic coupling in short SJJ's. However, magnetic
coupling in moderately long SJJ's strongly reduces metastability
of $I_c(\Phi_e)$. An additional large scale modulation of
$I_c(\Phi_e)$ in moderately long SJJ's SQUID is due to flux
quantization inside SJJ's.

In order to eliminate differences caused by the finite junction
length effect, a short SJJ SQUID with a smaller SQUID loop
$L_0=2\lambda_J$ was also simulated. Results are shown by open
circles in Fig. 2 b) along with the data for a moderately long SJJ
SQUID from Fig. 2 a). Now finite length effects are similar for
both short and long SJJ SQUID's. However, metastability is much
more pronounced for a short SJJ SQUID, while it is strongly
suppressed in a moderately long SJJ SQUID.

Next we consider the influence of critical current nonuniformity
and the dependence on the coupling strength. In Fig. 3 a) and b)
current-flux characteristics of moderately ($S=0.3$) and strongly
($S=0.5$) coupled SJJ SQUID's, respectively, are shown for
identical $J_c$'s (solid symbols) and for a 20 \% spread in
$J_c$'s (open symbols). In Fig. 3 a) magnetic coupling is reduced
due to a larger thickness of middle electrodes, see Eq.(2),
$d_{2,3}=0.1\lambda_J=\lambda_S$, other parameters are the same as
in Fig. 2. In the case of nonuniform junctions, $J_{c2}=1.1
J_{c1}$, $J_{c3}=1.2 J_{c1}$.

From Fig. 3 it is seen that $I_c(\Phi_e)$ of a moderately coupled
SJJ SQUID is metastable and the performance is further
deteriorated by the spread in $J_c$'s. On the contrary, strongly
coupled SJJ SQUID is more stable and less susceptible to a spread
in junction parameters.

The increased stability of strongly coupled SJJ SQUIDs can be
understood in terms of phase-locking of junctions in the stack.
Indeed, the energy difference between states is increased
approximately proportional to the number of phase-locked
junctions. This was confirmed experimentally by analyzing the
"escape rate" in HTSC IJJ's \cite{Mros}. Another important
advantage is that phase-locked junctions switch simultaneously
into the resistive state \cite{Mros}. This creates a possibility
for compensation of a reduced current-flux modulation in the
multi-junction SQUID by an increased voltage-flux modulation due
to phase locking of junctions in the stack.

\section{Conclusions}

In conclusion, operation of a stacked Josephson junction SQUID is
studied numerically. I demonstrate that the coupling between
stacked Josephson junctions can improve operation of a
multi-junction SQUID:

(i) Metastability of $I_c(\Phi)$ is reduced in SJJ's SQUID due to
magnetic coupling of junctions.

(ii) Strongly coupled SJJ's SQUID becomes more stable and less
susceptible to a spread in junction parameters.

(iii) Small $I_c(\Phi)$ modulation can be compensated by an
increased voltage-flux response if junctions in the stack are
phase-locked.

Finally, some remarks about the optimum SQUID design: Magnetic
coupling is effective in long SJJ's. However, SJJ's should not be
very long, otherwise multiple quasi-equilibrium fluxon modes in
SJJ's would strongly complicate flux penetration into the SQUID
loop \cite{Modes,Compar,Review}. The best performance is expected
for moderately long SJJ ($L_S \sim 4 \lambda_J \sim 2-3 \mu m$ for
Bi-2212). In order to achieve reasonable $I_c(\Phi)$ modulation
the number of SJJ's should be kept small (less than ten). This is
difficult for HTSC SQUID with the geometry shown in Fig. 1. The
solution would be to place the SQUID loop in the $ab$-plane and
form stacks of intrinsic Josephson junctions by cutting trenches,
restricting current flow in the $ab$-plane.

\end{document}